# Single-domain imaging in topological insulator $Bi_2Te_3$ thin films

David H. Yi[1], and Deepti Jain[2]

[1] *Bridgewater-Raritan High School, Bridgewater, New Jersey 08807, USA*
[2] *Department of Physics and Astronomy, Rutgers, The State University of New Jersey, Piscataway, NJ 08854, USA.*

**Abstract**

Single crystalline materials, different from polycrystalline and twinning structures, are desired for investigating the intrinsic physical properties, as grain and twin boundaries often work as a source of artifacts. Bismuth chalcogenides, which are van der Waals materials notable as topological insulators, have attracted significant interest due to their rich physical properties. However, the formation of 60° twin domains is common in these materials. Here, we demonstrate single-domain bismuth chalcogenides. Using atomic force microscopy, we investigated the morphology of $Bi_2Se_3$ and $Bi_2Te_3$ grown on $Al_2O_3$. Despite lattice constants of $Bi_2Se_3$ and $Al_2O_3$ substrates being well matched with hybrid symmetry epitaxy, $Bi_2Se_3$ exhibited 60° twin boundaries across the surface. Interestingly, $Bi_2Te_3$ showed a single-domain feature across the 10 mm by 10 mm sample even with lattice mismatch. While further in-depth studies are required to understand this difference in the morphology between $Bi_2Se_3/Al_2O_3$ and $Bi_2Te_3/Al_2O_3$, we suggest that the formation of twin boundaries in bismuth chalcogenides is related to the interaction between quintuple layers across the van der Waals gap rather than strain or defects.

**Introduction**

Single crystal refers to a material where the crystal lattice maintains a uniform orientation across the entire structure. This is in contrast with polycrystalline materials, which consist of regions with varying crystallographic orientations separated by grain boundaries, as shown in Figs. 1a and 1b.[1] The formation of these grain boundaries can be attributed to factors such as atomic defects, local off-stoichiometries, multiple nucleation sites and strain.[1-3] In the growth of epitaxial thin films, twin boundaries, characterized by mirror symmetry (Fig. 1c and 1d), are the most prevalent and readily formed grain boundaries. Therefore, achieving thin films without twin boundaries, dubbed twinless, is challenging. Even though these boundaries can sometimes result in unexpected and peculiar physical properties, the absence of such boundaries is desirable because it provides a platform for the material's intrinsic properties to be studied. This is due to the tendency for the physical characteristics at the boundary to differ from those of the intrinsic properties, which introduces artifacts that can obscure the investigation of the material's inherent properties. As tools to investigate the twin boundaries, atomic force microscopy (AFM), which provides the surface morphology, and transmission electron microscopy (TEM), which visualizes the crystal structure through the cross-sectional view of growth direction, are most widely employed.

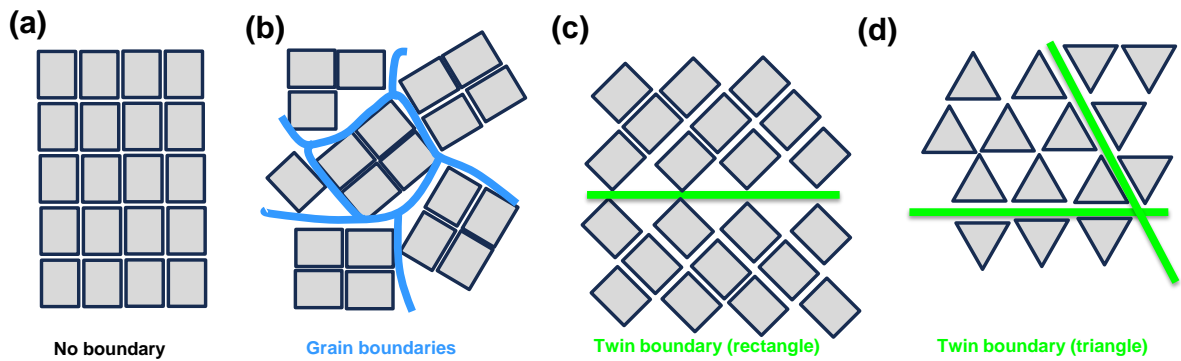

Figure 1: Type of boundaries of (a) single crystal, (b) polycrystals with blue-colored grain boundaries, and single crystals with green-colored twin domain boundaries for (c) rectangular and (d) triangular structures.

Here, we investigated the surface morphologies of molecular beam epitaxy (MBE) grown bismuth chalcogenides, $Bi_2X_3$, with X representing either Selenium, Se, or Tellurium, Te using AFM. These materials were chosen due to their hexagonal crystal structure, which facilitates the observation and analysis of domain orientations and the rich physical properties of topological insulators. For instance, $Bi_2Te_3$ and $Bi_2Se_3$ have been used as an ingredient in broad research fields such as thermoelectricity, nanoelectronics, optoelectronics, and spintronics.[4-7] Furthermore, its van der Waals nature allows it to be applied in heterostructure studies pursuing the various quantum phenomena like quantum anomalous Hall effect, Weyl semimetals, and Majorana physics[8-10]. In our study, we found that $Bi_2Se_3$ thin films show the randomly mixed 60° twin boundaries with a twin ratio of 1 to 1, consistent with previous reports.[11,12] On the other hand, in $Bi_2Te_3$ thin films grown on $Al_2O_3$, over 99 % of the areas have the same orientation. Based on this contrast in results, we speculate that the van der Waals (vdW) gap energy plays an essential role in the growth of twinless $Bi_2Te_3$ films. These findings will help further studies in bismuth chalcogenides and their applications.

**Results and discussions**

Figure 2a shows the crystal structure of $Bi_2X_3$ thin film with X = Se and Te. The crystal structure of the quintuple layer, QL, X-Bi-X-Bi-X, follows an ABC order, and the ABC sequence information transfers to the next QL across the van der Waals gap.

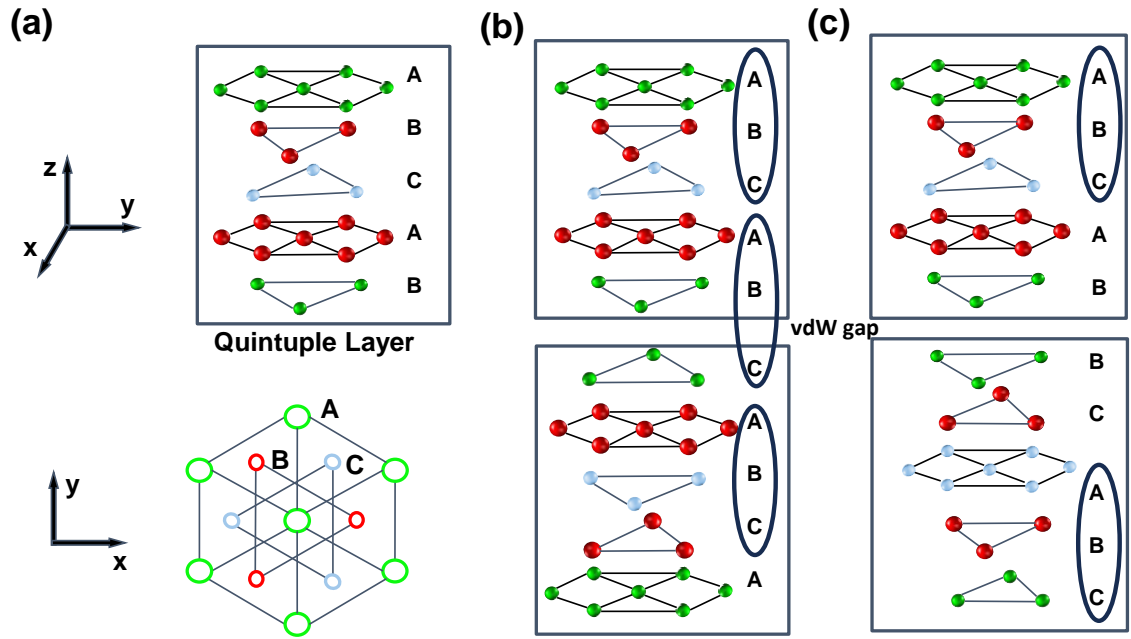

Figure 2. Crystal structure of $Bi_2X_3$ topological insulators. (a) top: X-Bi-X-Bi-X stacking of QL through growth direction. Bottom: top view of $Bi_2X_3$ QL. (b) shows the stacking of QLs with gaps in $Bi_2X_3$ material in ABC order through the van der Waals gap, resulting ABC/ABC. **(c)** The loss of the stacking pattern, ABC/ABB, induces the switching in orientation. Consequently, the 60°twin domains are across the thin film.

This implies that the upper QL has an ABCAB stacking sequence, and the next QL starts from C and keeps the ABC order. Therefore, the stacked two neighboring QLs follow ABCAB/CABCA sequence (Fig.2b). However, the loss of sequence across the van der Waals gap between neighboring QLs, ABCAB/BCABC (Fig. 2c), can also occur.

Figure 3a shows the AFM image of $Bi_2Se_3$ film taken 10 μm by 10 μm scan area. The average one-side length of a triangular pyramidal structure is 1.2 μm. The blue and red triangles show the 60° twin domains.

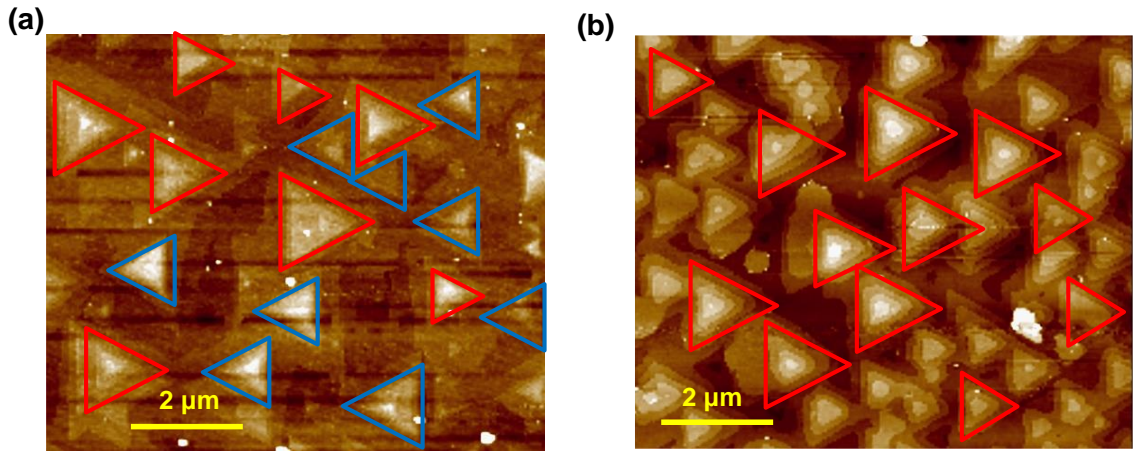

Figure 3 10μm by 10μm AFM image of (a) $Bi_2Se_3$ and (b) $Bi_2Te_3$. Red and blue triangles indicate the 60° twin domains in grown film, forming twin boundaries across the sample.

The number of blue and red triangles are almost the same, indicating that twin-domain boundaries are prevalent in this thin film. The crystal structure of QL, Se-Bi-Se-Bi-Se, follows an ABC order. However, the ABC order information does not transfer to the next QL if the sequence information is lost across the van der Waals gap between neighboring QLs. This implies that even though both upper and lower QLs follow an ABCAB stacking sequence, the ABC stacking sequence is broken at the interface of QLs. As shown in Fig. 2c, the lower QL starts from B instead of C and keeps the ABC order. Therefore, the stacked two neighboring QLs of ABCAB/BCABC tend to form twin boundaries that are oriented at 60° relative to each other (Fig. 3a). This 60° twin boundary in hexagonal thin films is the most common planar defect resulting from stacking faults during growth.[11,12] The stacking fault emerges when the atomic layers lose the sequence, likely caused by chemical defects, doping, and mechanical strains.

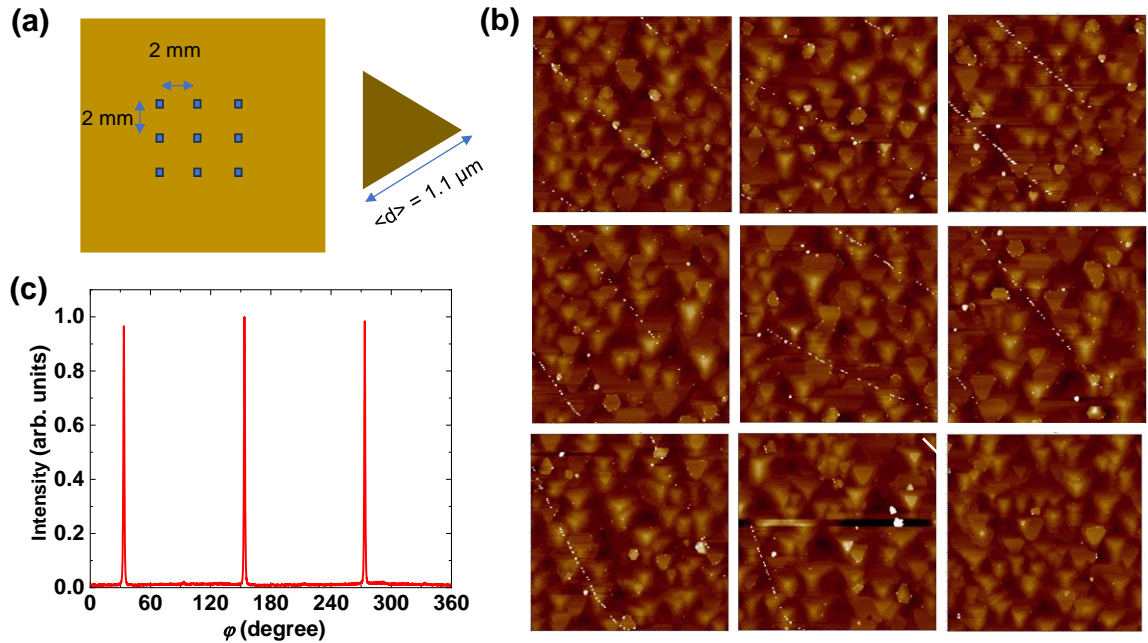

Figure 4. AFM image of $Bi_2Te_3$ (a) Schematic shows the nine scanned areas with a 2 mm distance on 10 mm by 10 mm grown film, (b) AFM image in nine different areas. Over 99 percent of triangles are aligned in the same direction, with an average triangle size of 1.1 µm. (c) In-plane XRD $\varphi$ scan of $Bi_2Te_3$.

However, when we scanned the morphology of $Bi_2Te_3$ thin film with AFM at a 10µm by 10 µm scale, we observed only one-directional triangles without any 60° twin domains (Fig. 3b). This implies that the arrangement of the basal planes in $Bi_2Te_3$ follows a sequence of ABC order and maintains this across the QLs. To confirm the twinless boundaries in $Bi_2Te_3$, we scanned nine different areas with 2 mm distances of 10 mm by 10 mm grown film, as shown in Fig. 4. Surprisingly, over 99 % of triangles in nine different areas align in one direction. There was an almost entirely uniform structure of the triangle pointing to the right-hand side, with a few inconsistent triangles pointing to the left-hand side. After testing several different $Bi_2Te_3$ samples, we confirmed that single-domain structures remained consistent. The x-ray diffraction azimuthal, $\varphi$, scan of $Bi_2Te_3$ thin film is another evidence of the single domain feature of the grown film. Each peak has a 120-degree spacing instead of 60°, confirming the twinless behaviour in the scanned sample.

**Discussions**

Detecting twin boundaries in $Bi_2Se_3$ but not in $Bi_2Te_3$ is fascinating due to their classification in the same chalcopyrite group and identical layered structures.

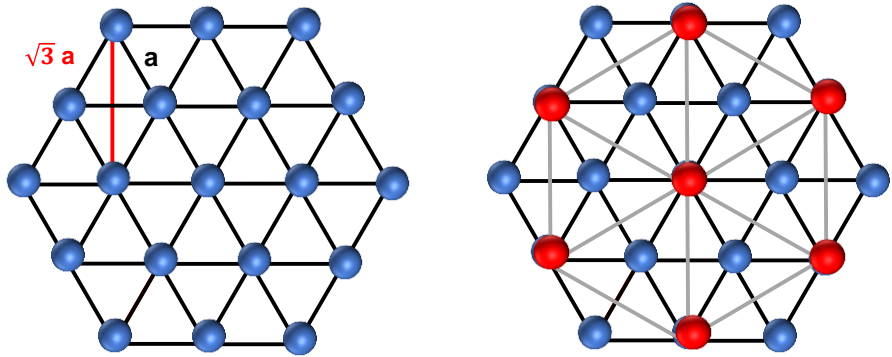

Figure 5. Schematic diagram of lattice match between two different materials with $2/\sqrt{3}$ relation.

Initially, we expected that there would be more chance to get the twinless in $Bi_2Se_3$ than in $Bi_2Te_3$ due to a hybrid symmetry epitaxy between $Al_2O_3$ and $Bi_2Se_3$. The hybrid symmetry epitaxy denotes that, even if there is a significant difference in lattice constants between neighboring materials, the film can grow epitaxially if there is a doubling or $3^{1/2}$ relationship between lattice constants of the growing material and substrate underneath, as shown in Fig. 5. The lattice constant of $Bi_2Se_3$ is 0.414 nm and that of $Al_2O_3$ is 0.476 nm which results in a negligible lattice mismatch of 0.4 %, if we consider the hybrid symmetry. On the other hand, the lattice constant of $Bi_2Te_3$ is 0.438 nm, and the compressive strain is over 6 %, even in the hybrid symmetry case. Due to this, we expected to observe the twin domains in $Bi_2Te_3$ rather than $Bi_2Se_3$ because of the strain and strain-induced defects in $Bi_2Te_3$ thin film.

Instead, this contrasting behavior can be traced to a nuanced difference in their atomic interactions and bonding attributes. One likely explanation is that the discrepancy stems from the distinct strengths of vdW interactions between the layers in these compounds. In $Bi_2Te_3$, the Tellurium-Tellurium (Te-Te) can be assumed to have stronger vdW bonds than the Selenium-Selenium (Se-Se) vdW bonds found in $Bi_2Se_3$. This enhanced Te-Te interaction may lead to a more stable and consistent layer stacking in $Bi_2Te_3$ thin film, thereby minimizing the formation of twin boundaries. On the other hand, the comparatively weaker Se-Se vdW bonding in $Bi_2Se_3$ may render its crystal structure more vulnerable to disruptions during growth. Such disruptions could facilitate the creation of twin boundaries, which occur when a 60° rotation misaligns the layer stacking sequence. Further, the differences in atomic size and electronegativity between Se and Te may also play a role in influencing the dynamics of crystal growth. Te atom has a larger size and less electronegative than Se atoms, potentially leading to variations in bonding and alignment during the crystallization process. Such variations may explain the discrepancies in twin boundary formation between $Bi_2Se_3$ and $Bi_2Te_3$.

**Conclusion**

We prepared $Bi_2Se_3$ thin films grown on $Al_2O_3$ to synthesize single-domain thin films based on the assumption that hybrid symmetry epitaxy, lattice match between $Bi_2Se_3$ and $Al_2O_3$ with $2/3^{1/2}$ relation and 0.4 % lattice match, would foster a single-domain thin film. As a comparison, $Bi_2Te_3$ thin films were also prepared, which have a 6 % lattice mismatch in hybrid symmetry. We investigated the morphology of $Bi_2Se_3$ thin film using the AFM and found that $Bi_2Se_3$ thin film has 60° twin boundaries across the surface, consistent with previous reports.[11,12] Surprisingly, scanned $Bi_2Te_3$ images show that only unidirectionally aligned triangles exist in 10 mm by 10 mm samples. The XRD azimuthal scan shows 120-degree spaces between each peak, confirming the twinless $Bi_2Te_3$ thin films grown on $Al_2O_3$ substrates. Even though more research is required to conclude, we suggest that the occurrence of twin boundaries in $Bi_2Se_3$ and their absence in $Bi_2Te_3$, is related to the interaction between QLs across the vdW gap rather than strain or defects. Gaining insights into these subtle distinctions is essential for optimizing the properties of these materials for targeted applications, especially in areas like thermoelectrics and spintronics, where crystal structure plays a crucial role in determining material performance.

**Materials and methods**
***Thin-film growth***: We prepared $Bi_2Te_3$ and $Bi_2Se_3$ films on $10\times10\times0.5$ mm$^3$ $Al_2O_3$ (0001) using a custom-built MBE system (SVTA) with a base pressure of ~$10^{-10}$ Torr. $Al_2O_3$ substrates were cleaned ex-situ by UV-generated ozone and in situ by heating to 750 °C under oxygen pressure of $5 \times 10^{-7}$ Torr. High purities of Bi, Te, and Se were thermally evaporated, using Knudsen diffusion cells for the film growth. All the source fluxes were calibrated *in-situ* by quartz crystal microbalance and *ex-situ* by Rutherford backscattering spectroscopy.

*AFM measurement*: 10µm by 10µm images around the sample have been scanned using aa tapping mode of AFM (NT-MDT Solver Nano.) with a resonance frequency of 213.870 kHz.

## Acknowledgements

We would like to thank Prof. S. Oh for allowing us to use the Topological Quantum Matter Engineering Lab facilities and for the helpful discussions.